\documentclass[preprint,aps]{revtex4}

\usepackage{graphicx}
\usepackage{dcolumn}
\usepackage{bm}

\begin{document}

	
\title{Lightning Strikes and Attribution of Climatic Change }

\author{Anthony J. Webster}

\email{anthony.webster@st-hughs.ox.ac.uk}

\date{\today}

\begin{abstract}
Using lightning strikes as an example, two possible schemes are
discussed for the attribution of changes in event frequency to
climate change, and estimating the cost associated with them.  
The schemes determine the fraction of events that should be attributed
to climatic change, and the fraction that should be attributed to
natural chance.   
They both allow for the expected increase in claims and the
fluctuations about this expected value.   
Importantly, the attribution fraction proposed in the second of these
schemes is necessarily different to that found in epidemiological
studies. 
This ensures that the statistically expected fraction of attributed
claims is correctly equal to the expected increase in claims. 
The analysis of lightning data highlights two particular difficulties
with data-driven, as opposed to modeled, attribution studies. 
The first is the possibility of unknown ``confounding'' variables that
can influence the strike frequency. 
This is partly accounted for here by considering the influence of
temperature changes within a given month, so as to standardise the
data to allow for cyclical climatic influences. 
The second is the possibility suggested by the data presented here,
that climate change may lead to qualitatively different climate
patterns, with a different relationship between e.g. strike frequency
and temperature. 
\end{abstract}

\maketitle

\section{Introduction}

Lightning strikes and the damage caused by them are generally expected
to increase as global temperatures rise 
\cite{Mills,Williams,Price1,Price2,Price3,Michalon,Toumi,Romps}.  
The mechanisms behind this are being determined 
\cite{Williams,Romps}, and despite changes in strike-frequency being
dependent on geographical location, the observed trend is very clear
\cite{Mills,Williams,Price1,Price2,Price3,Michalon,Toumi,Romps}.  
Consequently, lightning strikes provide an example where the
statistical connection between insurance claims, lightning strikes,
and temperature changes is increasingly well understood.   
A well known example is the Hartford Insurance
Co. lightning strike data presented in Ref. \cite{Mills}, that shows a 
very strong correlation between claim frequency and temperature. 
However, it is difficult to determine whether the relationship between
number of claims and temperature is a causal one, with an increased
temperature causing an increased number of strikes (and claims), or a
co-incidental correlation, with lightning strikes tending to occur
during the summer period for example.  
This is one of the questions that are considered here, as we explore 
the effectiveness of this and similar data, for lightning-strike
attribution studies.  
Presuming that a quantitative relationship can be determined between
climate change (through the proxy of temperature in this example), and
strike frequency, then in principle it becomes possible for changes in
the number (and cost) of insurance claims, to be paid for
by the polluter through a carbon tax \cite{CTax0,CTax1,NYTimes} or an
insurance-led levy \cite{WebsterClarke,Clarke}. 
For this to happen an additional difficulty needs to be overcome - to
devise a mechanism that fairly attributes (e.g.) insurance claim costs
between climatic change and natural statistical chance 
\cite{Pielke,Otto,Hannant}. 
This is considered in Section \ref{s2}. 
In section \ref{lstrikes} we will consider lightning strikes 
across the contiguous United States, and the extent to which
monthly-averaged data such as that presented in Ref. \cite{Mills} can
be used for attributing changes in lightning strike frequency to
climate change.  
This raises two interesting concerns for data-driven studies. 
The first is the influence of ``confounding'' variables that are not
included in the study, but also influence e.g. strike frequency. 
The second is the possibility of qualitative changes in climate
patterns making data-driven climate models unreliable. 
Section \ref{s2} considers two schemes for splitting the number of
strikes so that one fraction is attributed to climate change and the 
other to natural statistical chance. 
The lightning strike data is used to give an example of the proposed
attribution schemes in Section \ref{Ex.Att}, estimating the number of
strikes that should be attributed to climate change. 

\section{Lightning strikes}\label{lstrikes}

\begin{figure}
\centering
\includegraphics[width=0.8\linewidth]{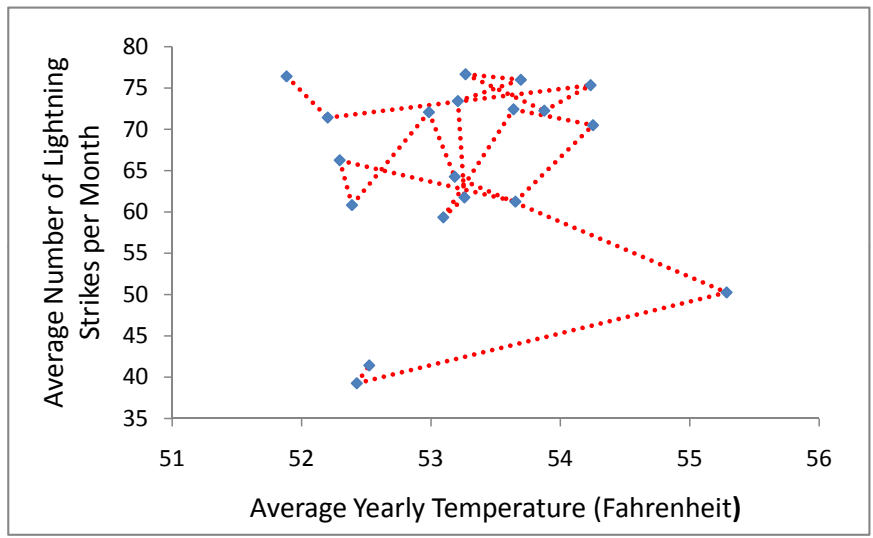} 
\caption{The average number of lightning strikes per month in the
  United States is plotted (vertical axis), versus the area-weighted
  yearly-averaged temperature. 
  The first data point is for 1996 in the top left of the plot, 
  the last is for 2014 near the bottom of the plot, and adjacent years
  are connected by a dashed line. 
  The majority of points for 1996-2011 are clustered
  around 60-75 strikes per month at a temperature of 52-54 degrees
  Fahrenheit, but the last three years 2012-2014 appear to be separated
  from the group. 
  We do not know whether this is due to changes in reporting, or if it 
  indicates a qualitative change in climate behaviour. 
}
\label{YearAv}
\end{figure}

Here we reconsider the relationship between reported lightning strike
damage and monthly averaged U.S. temperatures reported in
Ref. \cite{Mills}, using an enlarged database from 1996-2014, provided
by the National Climatic Data Centre (NCDC). 
The temperature data is from the GHCN monthly climate indices for the
area-weighted average temperature of the contiguous United States
\cite{Menne}.  
The lightning strike data is from the National Centre for
Environmental Information's Storm Events Database, that records
lightning strikes that result in personal injuries and damage. 
In principle this includes strikes in Alaska, the Gulf of Mexico,
Hawaii, and Hawaii waters, but these are comparatively small in
number, and can be neglected when compared to the numbers recorded for
the contiguous United States.  
We start by using this data to compare the average monthly strike rate
and the year-averaged temperature (figure \ref{YearAv}). 
Most of the data is grouped around 60-80 strikes per month with 
yearly-averaged temperatures of 52-54 degrees Fahrenheit, but
the last 3 years 2012-2014 are clearly separated from this group.  
We do not know whether this is due to changes in the reporting of
lightning strike events, whether it reflects a solar or other natural
climatic cycle, or whether
it is a symptom of a qualitative change in the climate.  
A straight line fit to figure \ref{YearAv} determines an increase
in strike frequency that equates to approximately 2.9\% per degree
Celcius, or 3.0\% per degree Celcius if the last three years 2012-2014
are omitted. 
These estimates are lower than the estimate of 12$\pm$5\% per degree
Celcius reported 
recently \cite{Romps}, but not necessarily inconsistent with it.
For example, if the strike frequency increases faster than
linearly with temperature then the average number of strikes is likely
to be greater than the number of strikes expected at the average
temperature (or in mathematical notation $E[N(T)] \geq N(E[T])$, which
is Jensen's inequality \cite{Jensen}).    
In practice this means that if the expected strike rate is calculated 
with measurements that are averaged over shorter monthly time-scales,
then we expect to find a higher estimate than that from figure
\ref{YearAv}.  
(In mathematical notation, if we sum over the months $i$ from $1$ to
$12$ then provided $N(T)$ is a convex function, then,   
$(1/12)\sum_{i=1}^{12} N(T_i) \geq N( (1/12)\sum_{i=1}^{12} T_i$, and
the total number of strikes in a year has 
$\sum_{i=1}^{12} N(T_i) \geq 12 \times N((1/12)\sum_{i=1}^{12} T_i )$)   
This is one simple way in which an estimate based on yearly averages
can be too low, others are discussed later, but it is related to a
more serious concern that is discussed next.   

\begin{figure}
\centering
\includegraphics[width=0.9\linewidth]{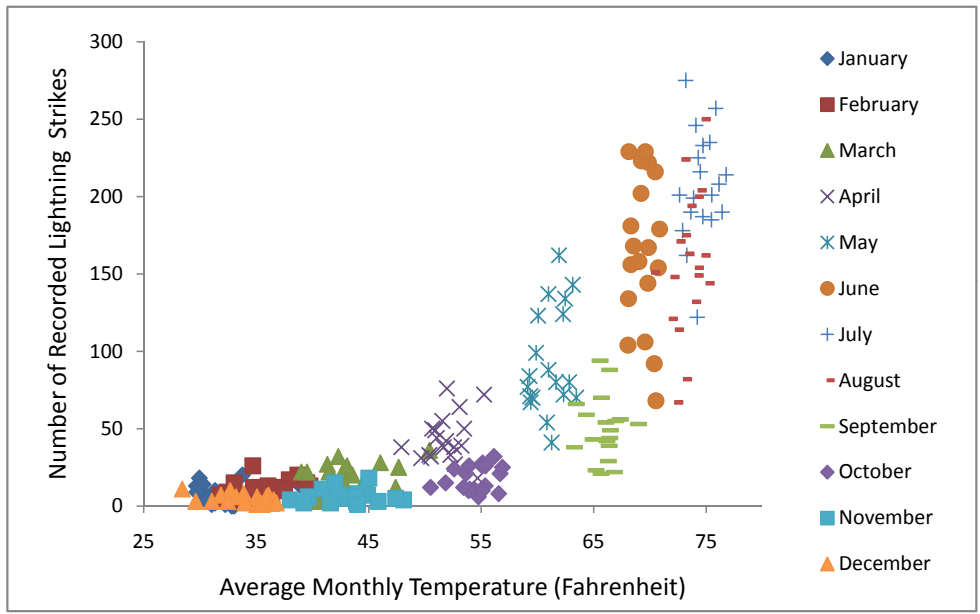}
\caption{For all years 1996-2014, the number of lightning strikes in a
month is plotted (vertical axis), versus the average monthly
temperature (horizontal axis). 
There is an approximately exponential increase in the rate of strikes
with temperature. 
However September is clearly hotter than April, but there are
generally less strikes in September, indicating that temperature
cannot be the only variable that influences the number of strikes. 
}
\label{fig:StrikesVsTemp-Monthly}
\end{figure}

\begin{figure}
\centering
\includegraphics[width=0.9\linewidth]{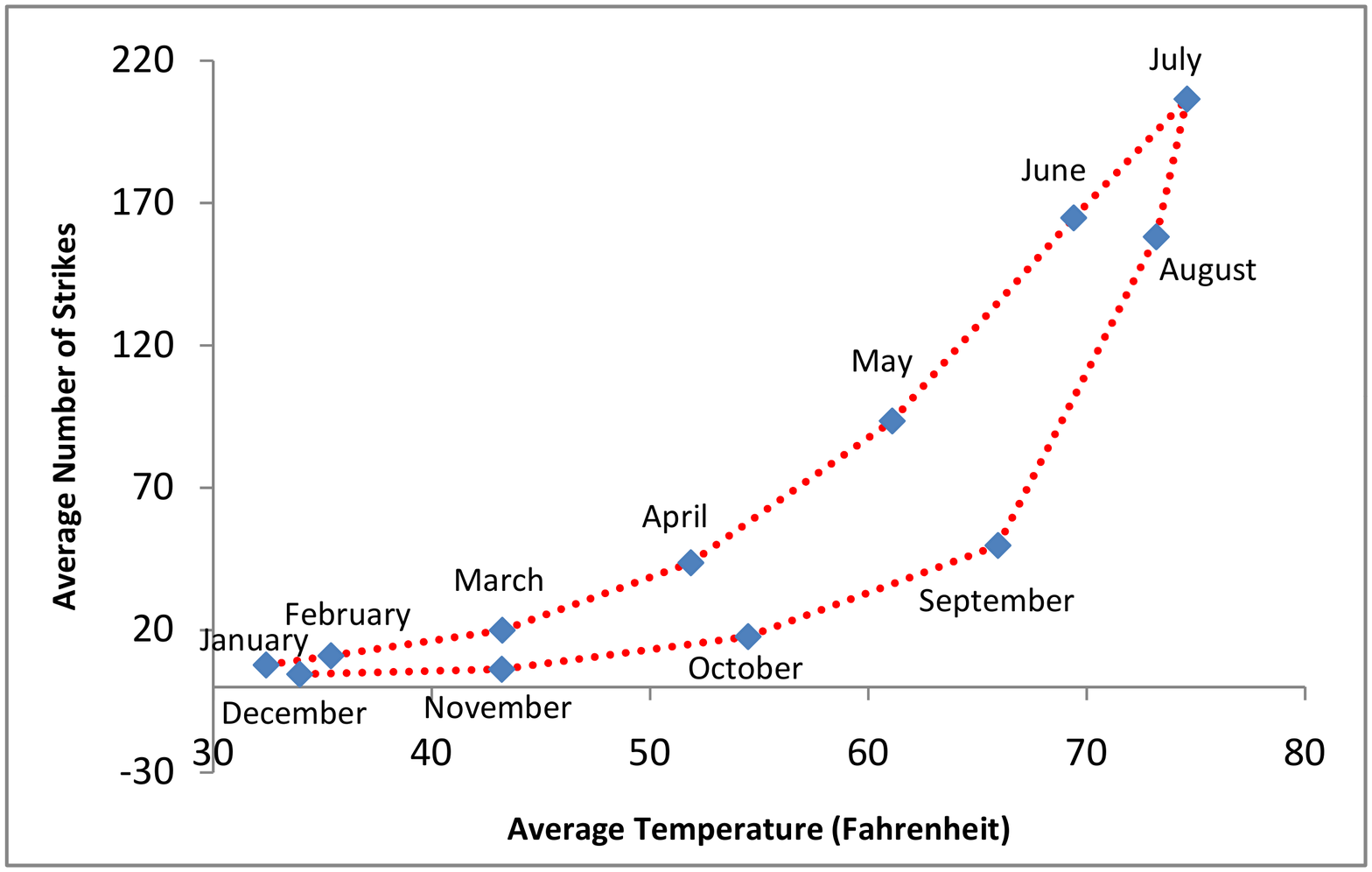}
\caption{The average number of lightning strikes in a month
  January-December (vertical axis), is plotted versus the average
  monthly temperature (horizontal axis).
  January is the leftmost data point, July is top-right, and December
  is bottom-left, with adjacent months connected by a dashed line. 
  There is a clear cyclical behaviour, indicating that initially the
  number of strikes increases more rapidly than temperature, and then
  subsequently falls more rapidly than temperature.
}
\label{CyclicNT}
\end{figure}

The Hartford Insurance Co. data \cite{Mills} suggests a strong almost
exponential relationship between the number of strikes and temperature. 
However, how can we be sure that the change in strikes is due to
changes in temperature, and not a coincidence, with the conditions
when strikes are more common being by chance at times of year
when the temperature is hotter? 
To explore this possibility we plot the number of strikes versus
average monthly temperatures, but grouped so that we know which
strikes arise from a given month.  
If you compare April with September for example, it is clear that
September has a higher average temperature than April, but the typical
number of strikes is lower.  
The conclusion is that temperature is not the sole factor determining
rate of strikes, there are other climate factors that must also be
accounted for.  
This becomes particularly clear if we plot the average number of
strikes in a given month versus the average temperature of that month
(figure \ref{CyclicNT}).  
This shows a clear cyclical behaviour with a higher rate of strikes at
a given temperature in the spring than in the autumn.  
The cyclical plot is because the rate of increase in strike rate early
in the year is faster than increases in temperature, and the rate of
reduction in strike rate later in the year is slower than the rate of
reduction in temperature.  
One interpretation is that the lightning strike rate has a more
immediate response to an increase in heat (energy) being input into
the climate system, whereas the temperature is slower to respond due
to the large thermal inertia (specific heat capacity) of water and
land mass. 
To help separate the influence of other possibly cyclical climatic
factors on strike rate from the influence of temperature, as far as is
possible with monthly-averaged data sets, we consider the relationship
between strike rate and temperature within each individual month.  
This approach is analogous to separating car drivers into different
age groups when trying to assess the influence of wearing glasses on
accident risk - older drivers are more likely to wear glasses than
younger drivers, recording drivers age groups helps to standardise
\cite{EpiUn} the data so that the influence of age is separated from
the influence of wearing glasses.   
The data available for each month is only 1/12 of that from the entire
year, which increases the statistical variability of the data.  
Nonetheless, to make quantitative statements about the probability of
events we would like a probability density function (``pdf'') for the
monthly data for average temperature and number of strikes.  
A simple Bivariate probability density function (pdf) is used to fit
the pairs of monthly data points $(N_i,T_i)$, with,  
\begin{equation}
P(N,T)= \frac{1}{2\pi \sigma_N \sigma_T}\frac{1}{\sqrt{1-\rho^2}}
\exp \left\{ - \frac{ \left[
\left( \frac{N-\mu_N}{\sigma_N} \right)^2 + \left(
  \frac{T-\mu_T}{\sigma_T} \right)^2  
- 2 \rho \left( \frac{N-\mu_N}{\sigma_N} \right) \left(
  \frac{T-\mu_T}{\sigma_T} \right) \right] }{ 2\left( 1 - \rho^2
\right) }\right\} 
\end{equation}
The parameters $\mu_N$, $\mu_T$, $\sigma_N$, $\sigma_T$, and $\rho$,
are here determined with a maximum-likelihood best fit \cite{Jensen}. 
The Bivariate pdf is a generalisation of the normal distribution to
two variables, and allows for a correlation between $N$ and $T$, that
is quantified by $\rho$.   
This gives a probability density function (pdf) for $N$ and $T$, $P(N,T)$. 
The Bivariate fit to the monthly data provides a pdf for pairs of
$(N,T)$, but we would like to know the pdf for the number of strikes
given the temperature, $P(N|T)$. 
Probability theory \cite{Jaynes} requires that $P(N,T)=P(N|T)P(T)$,
allowing us to obtain $P(N|T)$ from $P(N|T)=P(N,T)/P(T)$.  
If we approximate the pdf $P(T)$ for the (monthly-averaged)
temperature of a given month with a Normal distribution, then we can 
solve for $P(N|T)$ in terms of the fitting parameters for
$P(N,T)$, with, 
\begin{equation}\label{PNT}
P(N|T) = \frac{1}{\sqrt{2\pi}} \frac{1}{\sigma_N \sqrt{1- \rho^2}} 
\exp \left\{ - \frac{1}{2} \frac{1}{\sigma_N^2 (1-\rho^2)} 
\left[
N - \left( \bar{N}-\rho \bar{T} \frac{\sigma_N}{\sigma_T} \right) 
- \rho \frac{\sigma_N}{\sigma_T} T 
\right]^2 \right\}
\end{equation} 
Because $N$ and $T$ are averages (over a month), the choice of
Bivariate and Normal pdfs are suggested by the central limit theorem
\cite{Jaynes}, that with reasonable mathematical assumptions, requires
the pdf for an average to have a Normal distribution 
(or Bivariate when generalised to two variables). 
Notice the form of the expression within the square brackets of
Eq. \ref{PNT}, of $[N-a-bT]$, with 
\begin{equation}
a = \bar{N}-\rho \bar{T} \frac{\sigma_N}{\sigma_T}
\end{equation}
and 
\begin{equation}
b = \rho \frac{\sigma_N}{\sigma_T}
\end{equation}
The values of $a$ and $b$ are equivalent to those that would be
obtained from a least squares best fit for $N(T)$, but now we
additionally have an estimate for the standard deviation of
$\sigma_N(1-\rho^2)$, and more importantly a pdf for $P(N|T)$.  
The value of b determines how the expected number of claims will
increase with temperature in a given month. 
If we average $b$ over all the months, we obtain an estimate of
a 5.6\% increase in claims per degree C.
It is not surprising that the monthly-averaged estimate is different
to the yearly-averaged one discussed in Section \ref{lstrikes}. 
Using the example discussed in Section \ref{lstrikes}, if the
relationship between strike rate and temperature were convex then an
increased estimate is what we would expect from Jensen's inequality
\cite{Jensen}, and the estimate would be expected to be increased
further if we used weekly- or daily-averaged data.  
However the situation is more complicated, as is discussed shortly. 
The estimate of 5.6\% is made using recorded observations of lightning
strikes on the ground, and is closer to the value of 12$\pm$5\% estimated
from satellite data and climate modeling in Ref. \cite{Romps} than
the estimate using the yearly-averaged data in figure \ref{YearAv}.   
According to the data, the change in strike rate with temperature
varies considerably for different months. 
The largest increases in strike rate are for August (averaging 158
recorded strikes per month, but estimated to increase by 25\% per
degree Celcius), and May (averaging 93 recorded strikes per month, but
estimated to increase by 13\% per degree Celcius).
In contrast, for June there is an expected {\sl decrease} in strike
rate (averaging 165 recorded strikes per month, with a -9\% expected
change in strike frequency per degree Celcius).  
July presently has the most recorded strikes (averaging 207 per
month), and these are expected to increase by about 7\% per degree
Celcius.  
The average monthly changes in temperature from those in the 20th
century are not uniform either. 
The data used here gives the average increase in temperature during
the period 1996-2014 (across the spatially averaged contiguous United
States), as 0.67 degrees Celcius (1.21 degrees Fahrenheit), from those
in the 20th century.  
However the increase in monthly temperatures are greatest for the
months from November to March, with an average increase of 0.90
degrees Celcius (1.63 degrees Fahrenheit), but smaller for the months
from April to September whose average increase is 0.53 degrees Celcius
(0.90 degrees Fahrenheit).  
Consequently the change in strikes with temperature will not simply be
5.6\% per degree Celcius, because the temperature changes expected in
each individual month can be very different to the average across all
months.  
An average increase in temperature by 0.67 degrees Celcius from the
20th century average would give an estimated increase in strikes of
0.67$\times$5.6=3.8\%. 
In contrast, if we use the 20th century average monthly temperatures
to estimate the equivalent number of strikes for 20th century climate, 
then we estimate a reduction in the
average number of strikes per year from 784 during 1996-2014 to 758.  
This suggests that the number of strikes has increased by 3.4\% since
the 20th century (100$\times$25.7/758.1).  
This is slightly less than the 3.8\% increase that was estimated using
the average value of $b$ and the average change in temperature. 
It equates to a 5.1\% increase in strikes per degree Celcius increase
in {\sl average} temperature. 
Because this latter estimate more accurately accounts for non-uniform
changes in monthly temperatures, in principle it would be expected to
be more accurate than estimates using the average monthly response
(average $b$), to an average change in temperature. 
The average yearly standard deviation in the number of strikes can be
estimated from $\sum_{i=1}^{12} \sigma_{N}(i)\sqrt{1-\rho^2(i)}$, where
$\sigma_N(i)$ is the standard deviation of the number of strikes in
the $i$th month, and $\rho(i)$ is the correlation coefficient for the
$i$th month. 
Normalising this by the total average yearly number of strikes gives
an estimated variation of 29\%, which is considerably higher than the
expected increase in temperature per degree, independent of whether
an estimate of 5.6\% or 5.1\% are used. 

\section{Attribution of climate change}\label{s2}

The climate is only statistically predictable. 
The average expected temperature can be predicted, but the specific
temperature cannot.  
Consequently the average number of claims due to climate change are
in principle predictable, but not the actual number.  
Therefore if a carbon tax for example were used to pay for the
increase in claims, then an immediate question is whether it should
contribute to the expected increase in claims, or the actual number
that is observed? 
Two options that we will explore in more detail shortly are:
\begin{itemize}
	\item[(A)]{Pay a predetermined rate as a carbon tax
            \cite{CTax0,CTax1,NYTimes} or insurance levy
            \cite{WebsterClarke,Clarke}, possibly annually, that is
            based upon the {\it expected} number of claims.} 
	\item[(B)]{Pay for a fraction of the total claims that
            actually {\it occur}, with 
		that fraction determined in a way that reflects the relative
		likelihood for that number of claims occurring in the presence
		(and absence) of climate change.} 
\end{itemize}
The advantage of (B) is that the actual number of claims made are
subsidised, so the insurance industry does not profit in good years,
and is fully compensated in bad years.  
The disadvantage of (B) is that the subsidy paid to the insurance industry
(and claimed from e.g. carbon emitting industries), will fluctuate in
response to the natural statistical variability of the climate.  
Similarly, a disadvantage of (A) is that the subsidy paid or received
(some claims might reduce due to climate change), would only on
average be expected to match the actual changes in claim size;
although the uncertainty in the average change in claim size can be
estimated.  
An option (C) is to ``split the difference'', so that half the costs
are paid by method (A) and half by method (B), with the advantages and
disadvantages shared by both the insurance industry and whoever is
paying the subsidy.  
In practice this might involve the payer of the subsidy paying extra
to insure themselves against such cost increases.  
This suggests a simpler alternative that is similar to (A), whereby
the subsidy payer pays an additional premium to acknowledge that the
burden of uncertainty has been passed to the insurance companies.  
This situation is virtually equivalent to (C), but would operate like
a simple modification of (A), and is the simplest scheme that captures
the consequences of both the expected increase in claims and the extra
costs due to fluctuations of the actual claim sizes.  
The simplicity of this is attractive, but it does require the
insurance industry to be able to absorb any large fluctuations of the
claim sizes from their expected values, whereas with scheme (B) the
costs would be spread more broadly across both the insurance industry
and those contributing to a carbon tax \cite{CTax0,CTax1,NYTimes} (or
insurance levy \cite{WebsterClarke,Clarke}).   

\section{Attribution pricing mechanisms}\label{s3} 

Two simple attribution pricing mechanisms for schemes (A) and (B) are
described next. 
The main purpose is to show that simple, pragmatic, pricing schemes
can be formulated, and to highlight some of their properties, and some
properties required of them. 
A variety of alternative pricing schemes could no-doubt be proposed.  
For simplicity we discuss the total number of claims made, as
opposed to their total value, but the approach is easily modified. 

\subsection*{Scheme (A)}

\begin{figure}
	\centering
        \includegraphics[width=0.8\linewidth]{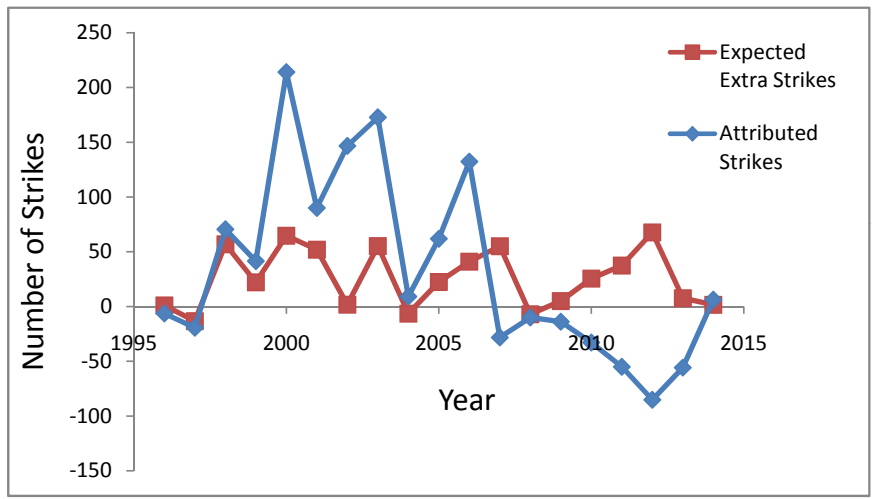}
	\caption{The attributed increase in lightning strikes compared
        to that expected for the 20th century are
        plotted (vertical axis), versus year (horizontal axis), for
        two attribution schemes (A) and (B). 
        The expected increase in strikes (``Expected Extra Strikes'') are
        calculated from Eqs. \ref{PNT} and \ref{AdN}, using the
        temperature data from 1996-2014 and the average 20th century
        temperature data. 
        The number of strikes attributed to climate change using
        scheme (B) (``Attributed Strikes''), are calculated from
        $N(1-\alpha)$ with $\alpha = p(N|T_0)/P(N|T)$ and using
        Eq. \ref{PNT} with 1996-2014 and 20th century average monthly
        temperatures, and the observed numbers of strikes $N$. 
        As expected, there are larger fluctuations with scheme (B)  
        because they are proportional to the actually
        observed numbers of strikes. 
        Because natural statistical variations can make the
        temperature less than 20th century averages, there are some
        years where the ``Attributed'' and ``Expected'' strikes are
        negative.  
        As discussed in Figure \ref{YearAv}, the years 2011-2014 have an
        unexpectedly low number lightning strikes. 
}
	\label{AttY}
\end{figure}

\begin{figure}
\centering
\includegraphics[width=1.0\linewidth]{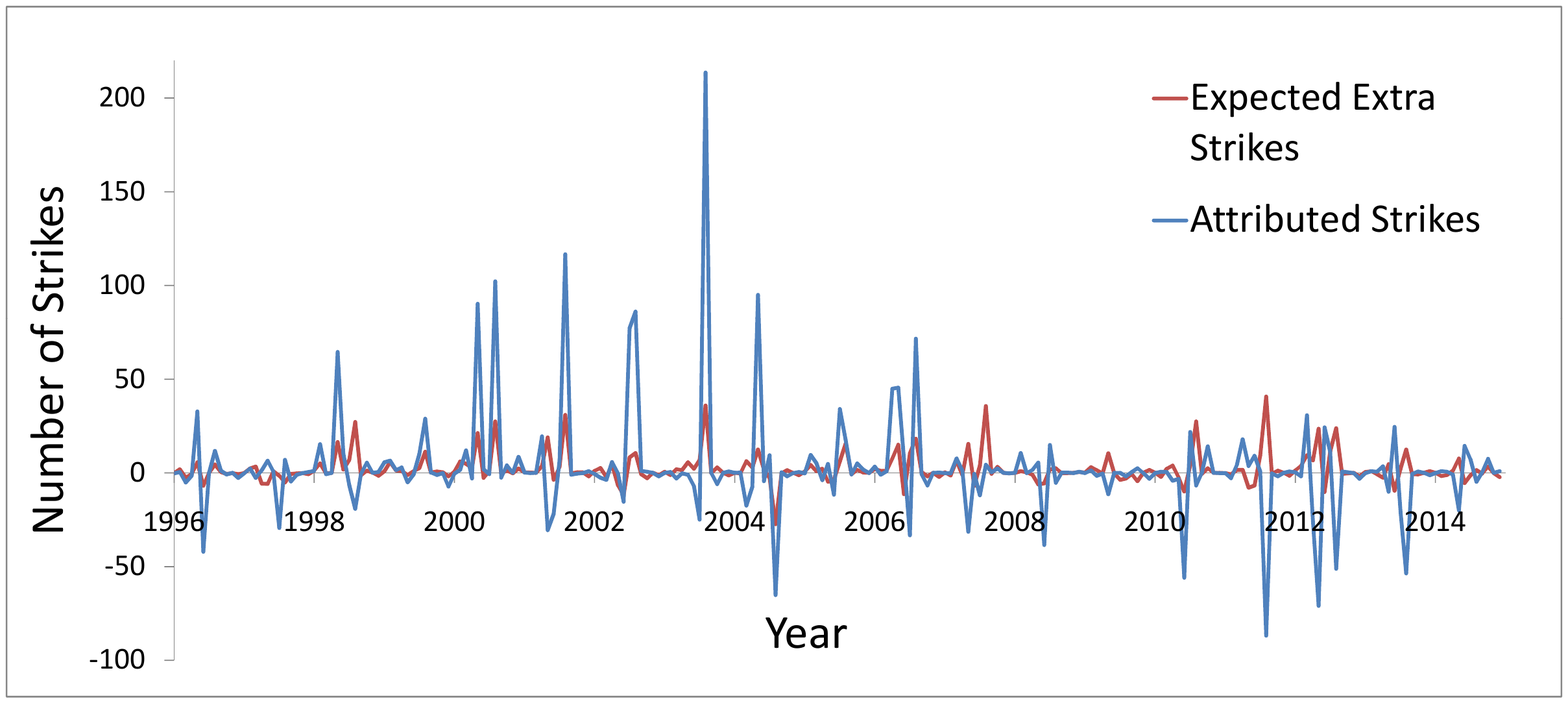}
\caption{Similarly to figure \ref{AttY} the attributed increase in
  lightning strikes compared to that expected for the 20th century are
  plotted (vertical axis), but now versus each month in 1996-2014
  (horizontal axis), for the two different attribution schemes (A) and (B).
 }
\label{AttM}
\end{figure}

Let $P(N|T)$ be the probability of $N$ claims given a change in
temperature from $T_0$ to $T$, with $T_0$ corresponding to the 20th
century average temperature for a particular month for example. 
With greater generality we could have used atmospheric carbon dioxide
levels in place of temperature $T$, but for clarity temperature is
used here.  
Then we can write $P(N|T)$ as, 
\begin{equation}\label{Pav}
P(N|T) = P(N|T_0) + \left[ P(N|T) - P(N|T_0) \right] 
\end{equation}
which is an exact identity. 
Multiplying (\ref{Pav}) by $N$ and integrating from $0$ to $\infty$
with respect to $N$, gives, 
\begin{equation}\label{AdN}
E\left[ N | T \right] = E \left[ N | T_0 \right] + \delta E[N|T] 
\end{equation} 
where the notation $E[N|T]$ denotes the expected value of N obtained
by integration, and $\delta E[N|T]\equiv \int_0^{\infty} dN [ N P(N|T) -
N P(N|T_0) ]$ denotes the change in the expected number of claims due
to a change in temperature from $T_0$ to $T$. 
In principle $\delta E[N|T]$ can be positive or negative, depending on
whether $P(N|T)$ is greater or less than $P(N|T_0)$, i.e. depending on
whether climate change makes a claim more or less likely.   
An advantage of this simple approach is that $\delta E(N|T)$ can be
calculated in advance using a climate model, and the estimate would
then be determined in advance and would not fluctuate, unlike the
actually observed values.  
The expected size of fluctuations about $\delta E(N|T)$ can also be 
estimated comparatively easily with varying degrees of sophistication.      
Alternatively, Eq. \ref{AdN} can be used with the observed temperature
data to give the expected number of extra strikes for the temperature
$T$, compared with the expected temperature $T_0$ for that month, this
is done later in figures \ref{AttY}-\ref{NatM}. 
Note that depending on the values of $T$ and $T_0$, $\delta E[N|T]$
can be either positive or negative; it is possible that climate change
can lead to less lightning strikes, certainly within individual
months.  
Another advantage of this approach are its mathematical similarities
to 
methodologies used in financial analysis, such as Modern Portfolio
Theory (MPT) \cite{MPT} and Credibility Theory \cite{Buhlmann},
suggesting that it could easily be incorporated into financial models.  

\subsection*{Scheme B}

Scheme B requires that the total cost of actual claims that occur is
paid partly from a carbon tax \cite{CTax0,CTax1,NYTimes} or an
insurance-led levy \cite{WebsterClarke,Clarke}, and partly by the
individual insurance companies.    
This needs to be done fairly, and must recognise that
statistically, any number of claims could in principle occur,
independent of whether climate change happens or not.  
It also needs to recognise that statistically at least, climate change
will modify the number of claims that are expected to occur, and it
should be able to estimate the increased (or decreased) proportion of
claims so that they can be correctly paid for by a carbon tax for example.  
If we divide (\ref{Pav}) by $P(N|T)$, and then multiply it by the
observed number of claims $N$ then we have, 
\begin{equation}\label{Nexact}
N = N \frac{P(N|T_0)}{P(N|T)} + N \left( 1 - \frac{P(N|T_0)}{P(N|T)} \right)
\end{equation}
This equation (\ref{Nexact}) is exact, and 
has the property of splitting the claims so that $N=N \alpha + N
(1-\alpha)$ with $\alpha = P(N|T_0)/P(N|T)$.  
When the expected number of claims $N$ is independent of climate
change with $P(N|T_0)=P(N|T)$, then $\alpha = 1$, a situation that
would correspond to zero subsidy being paid. 
However if the expected number of claims has dramatically increased
due to climate change, with $P(N|T)$ much larger than $P(N|T_0)$, then
$\alpha \rightarrow 0$, a situation corresponding to all the claims
being paid for by a carbon tax (or insurance levy).  
The split between the payments is determined by $\alpha =
P(N|T_0)/P(N|T)$, that is determined by a quantitative model for the
relative probability $\alpha$ of observing $N$ claims when climate
change is, or is not, present. 
In the example here $P(N|T)$ is estimated from the observed data for
lightning strikes, $T_0$ is the equivalent 20th century average
monthly temperature, and $T$ is the observed monthly average
temperature; but in principle there is no reason why a more
sophisticated climate model should not be used.   
In principle $P(N|T)$ and $P(N|T_0)$ can be determined from climate  
models, to give the relative probability of observing $N$ claims when
climate change is, or is not, present.   
Actuaries will be familiar with the form of Eq. \ref{Nexact}, because
it is the same as the widely used credibility theory estimators
\cite{Buhlmann}.  
The fraction $(1-\alpha)=\left( P(N|T) - P(N|T_0) \right)/P(N|T)$ will
also be familiar to epidemiologists, because it is very similar to the
``excess fraction'' \cite{ModEp} of extra incidences that result from
exposure to new climate conditions.  
This latter point suggests that similar schemes could be used for
other applications, such as quantifying the contribution an individual
may have to a team's performance for example.  
A difference between $(1-\alpha)$ and the excess fraction used in
epidemiological studies, is that here the ratio
$\alpha=(P(N|T_0)/P(N|T)$ is the inverse to what would usually 
define an excess fraction in epidemiology \cite{EpiUn,ModEp}. 
This is deliberate and the reason why is explained below. 
The excess fraction that we define considers
$P(N|T_0)/P(N|T)$, whereas the excess fraction used in epidemiological
studies would consider $P(N|T)/P(N|T_0)$ \cite{EpiUn,ModEp}. 
Why have we done this, and which is correct in this instance?
On average, we would like the expected value of the attributed
fraction of claims to equal the expected increase in claims as
estimated by scheme (A).   
Therefore given the temperature and climate conditions and the
probability density $P(N|T)$ associated with the temperature and
climate conditions at the present (or future time), then we would like
$E[N(1-\alpha)|T]$ to equal  
$E[N|T]-E[N|T_0]$. 
For the excess fraction that we define this is exactly the case, 
\begin{equation}
\begin{array}{ll}
E[N(1-\alpha)|T] 
&= \int_0^{\infty} N \left( 1 - \frac{P(N|T_0)}{P(N|T)} \right) 
P(N|T) dN 
\\
&= E[N|T] - E[N|T_0]
\end{array}
\end{equation} 
and the expected value of the fraction of claims attributed to climate
change $N(1-\alpha)$, at the present or future climate conditions with
temperature $T$, equals the expected increase in claims due to climate
change.   
The requirement of $E[N(1-\alpha)]=E[N|T]-E[N|T_0]$ is the reason why
the excess fraction that we have defined with $\alpha =
P(N|T_0)/P(N|T)$, is the correct one to use in this case, but is the
inverse of the excess fraction usually used in epidemiology studies that
would instead consider $P(N|T)/P(N|T_0)$ \cite{EpiUn,ModEp}.  
The separation of payments into those paid by the insurance company
($N\alpha$), and those paid by a climate change price ($N(1-\alpha)$),
has all the properties we would want of a pricing scheme: 
i) Equation (\ref{Nexact}) is exact, there will be no over (or under)
repayments in any given year,  
ii) The split in payments is determined in a quantitative way using
probability theory, 
iii) The possibility that any given claim could statistically have
occurred by chance is implicit, as is the recognition that the likely
claim size can be modified by climate change, 
iv) Its expected value is the same as the expected increase in claims
due to climate change, i.e. its average expected value is the same as
scheme (A),  
v) It is simple to understand, and intuitively fair.  

\section{Example: Attribution of U.S. Lighting Strikes}\label{Ex.Att}

\begin{figure}
	\centering
	\includegraphics[width=0.8\linewidth]{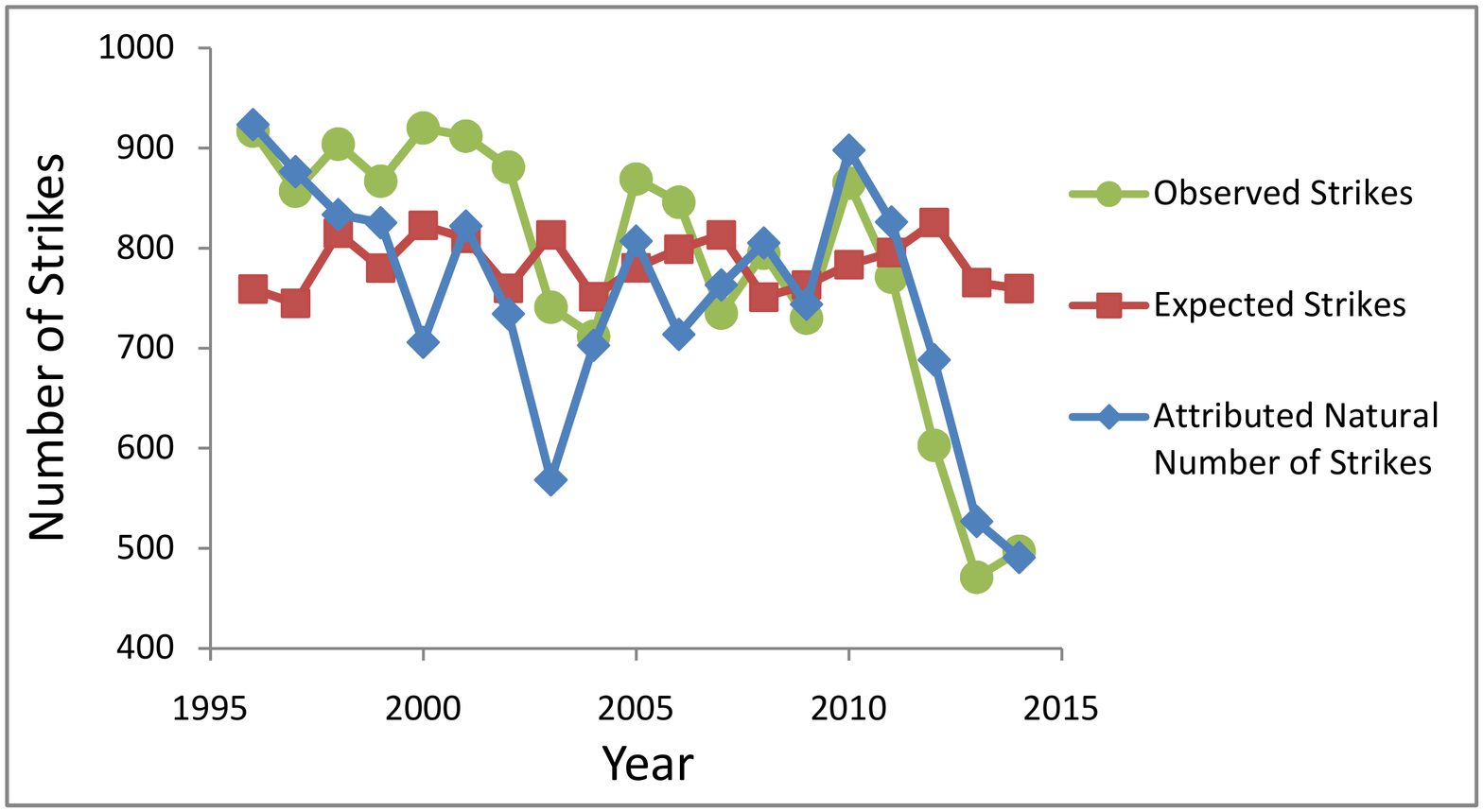}
	\caption{The observed lightning strikes are plotted (vertical
          axis) versus year, along with the expected number of strikes
          as calculated using the observed temperature data and
          Eq. \ref{PNT}, and the number of strikes that would be
          attributed as ``naturally'' occurring ($N \times P(N|T_0)/P(N|T))$.
        The different advantages of schemes (A) and (B) are clear -
        scheme (A) gives a less volatile estimate, but scheme (B)
        remains closer to the actually observed values.}
	\label{NatY}
\end{figure}

\begin{figure}
\centering
\includegraphics[width=1.0\linewidth]{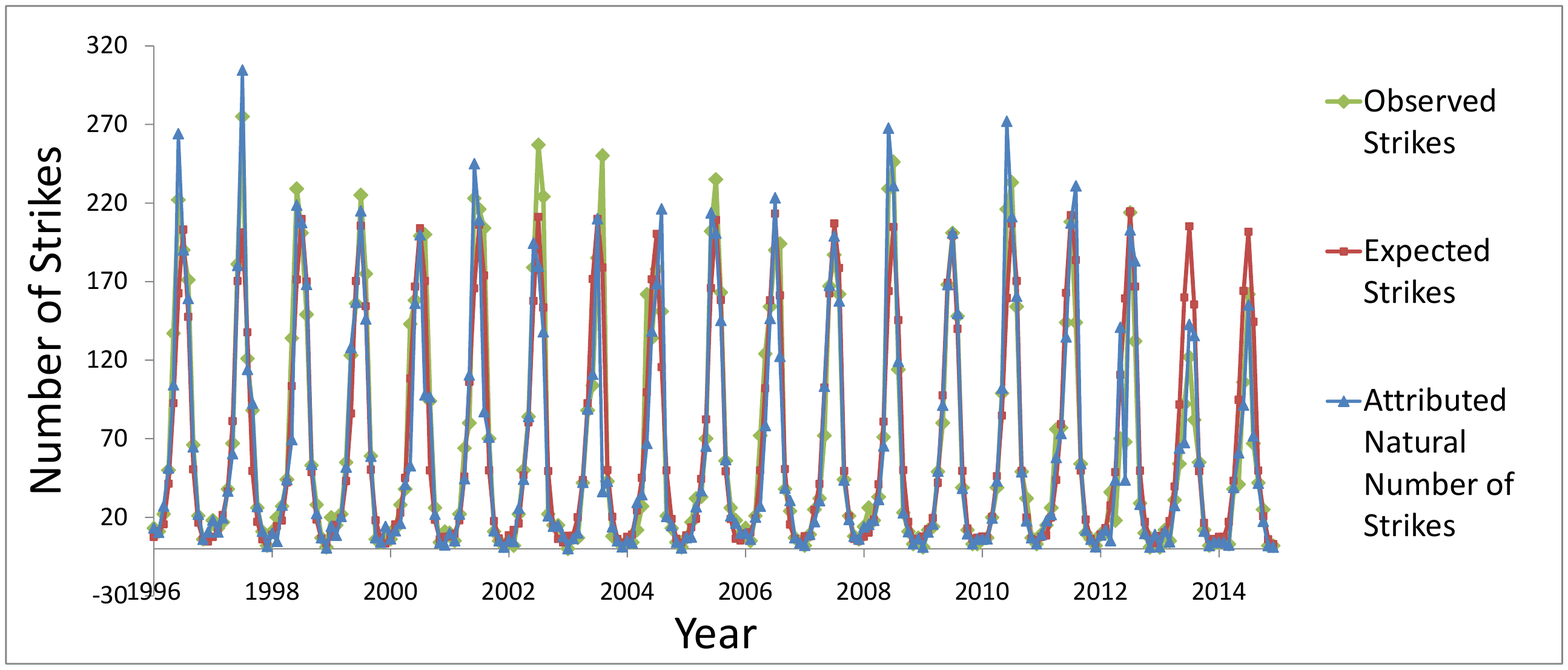}
\caption{Similarly to figure \ref{NatY} the observed lightning strikes
  are plotted (vertical axis), but now versus month in years
  1996-2014, along with the expected number of strikes as calculated
  using the observed temperature data and Eq. \ref{PNT}, and the
  number of strikes that would be attributed as ``naturally'' occurring
  ($N \times P(N|T_0)/P(N|T))$.
  The monthly data clearly shows the periodic variation in frequency
  of lightning strikes through the year from winter to summer, with
  the majority of strikes occurring in the warmer spring and summer
  months. 
} 
\label{NatM}
\end{figure}

As described in Section \ref{lstrikes}, we estimate an average
increase in strikes of 5.6\% per degree Celcius, or 5.1\% if the
estimate compares the observed number of strikes during 1996-2014 with
the estimated number of strikes for average 20th century monthly
temperatures.  
We can estimate the cost of increased strike frequency by estimating
the equivalent 20th century strike rate (estimated in Section
\ref{lstrikes} to be 758 strikes per year), and the damage cost
estimates recorded in the NCDC storm events database that give  
an average cost per claim for the 
period 1996-2014 of \$57800 (not adjusted for inflation). 
A 5.6\% increase per degree corresponds to an estimated cost of \$1.6M
for the 0.67 degrees Celcius increase in average temperature in
1996-2014 above the average 20th century temperature. 
At the present estimated warming rate of 0.19 degree C per decade, in 
ten years the costs are estimated to rise to roughly \$2.1M due to
climate change (no adjustments for inflation in these estimates).  
If the estimate of a 5.1\% increase in strikes per degree Celcius is 
used, then the estimated cost increases would become \$1.5M for
1996-2014, increasing to \$1.9M within ten years.  
Although the increase in (reported) costs are comparatively small
compared with the US economy, they correspond to an increased strike
frequency of approximately 4.8\% and 4.4\% (above those expected for
20th century temperatures).    
Section \ref{s3} discussed two alternative attribution pricing
mechanisms, scheme (A) that used the expected change in strike
frequency, and scheme (B) that attributes a fraction of the total
observed strikes to climate change. 
The number of strikes attributed to climate change is plotted for both
methods in figures \ref{AttY} and \ref{AttM}, that show the values
averaged over a year and for individual months respectively. 
As expected, the variation in attributed numbers of strikes is 
greater for scheme (B) than for (A), which would make it more
difficult to anticipate the costs using scheme (B). 
Another advantage of scheme (A) is that climate modeling could be
used to estimate or define the expected number of attributed strikes
in advance. 
However there are different benefits for scheme (B) over (A). 
Figures \ref{NatY} and \ref{NatM} plot the actual number of claims,
and the number of naturally expected claims given the observed
temperatures using schemes (A) and (B), averaged over a year and for
individual months respectively. 
The oscillations seen in figure \ref{NatM} are due to seasonal
variations in climate and temperature over a yearly cycle. 
The advantage of scheme (B) is easiest to see from figure \ref{NatY}.  
Whereas the expected number of naturally occurring strikes remains
comparatively stable, the attributed number of naturally expected
strikes remains similar to the number of observed strikes. 
This helps to avoid subsidising claims that do not occur, and to
ensure that if a large number of claims do occur, that they are fairly
compensated. 
Arguably scheme (B) more accurately reflects the actual costs (or
not), of climate change, but the costs are less predictable than for
scheme (A). 
This is particularly true at present, because the natural yearly
statistical variation in the number of strikes (estimated in Section
\ref{lstrikes} to be of order 29\%), is much greater than the expected
change in strike frequency due to climate change (the largest estimate
here is a 5.6\% increase in frequency). 
The average number of strikes per year attributed to climate change
during 1996-2014 by scheme A is 26, corresponding to a 3.4\% increase
compared with those expected for 20th century averages. 
This is equivalent to a 5.1\% increase in strikes per degree Celcius,
and is an identical calculation to the estimate in Section
\ref{lstrikes} that compared observed strike rates in 1996-2014 with
those expected for 20th century temperatures. 
The average number of strikes per year in 1996-2014 that are
attributed to climate change by scheme B is 34, corresponding to a
4.4\% increase above the number expected for average 20th century
temperatures.  
This equates to a 6.6\% increase per degree Celcius. 
If the last three years 2012-2014 are excluded, scheme B would
estimate that an average of 48 strikes per year should be attributed to
climate change, a 6.4\% increase, or 9.5\% increase per degree
Celcius. 
The estimates of scheme A remain fairly insensitive to whether the
last three years 2012-2014 are included or not. 
The estimates reflect the qualitative remarks made earlier, with a
greater volatility in estimates by scheme B than for scheme A. 
However scheme B more accurately reflects the number of strikes that
actually occur. 
As discussed in Section \ref{s3}, over a long enough time period, the
estimates using schemes A and B will converge to the same average value. 

\section{Conclusions}\label{conc}

Using monthly-averaged US temperature and lightning strike data as an
example, we have considered two different schemes for determining how
many lightning strikes should be attributed to climate change. 
Scheme (A) has the advantage of smaller fluctuations in the number of
attributed strikes than scheme (B), but scheme (B) has the advantage
of more accurately reflecting the actual number of observed strikes.
Therefore, if a carbon tax \cite{CTax0,CTax1,NYTimes} or insurance
levy \cite{WebsterClarke,Clarke} were used to pay for the
change in claim frequency (due to climate change), then scheme (B)
would avoid over or under compensating claims attributed to climate
change. 
The excess fraction used by scheme (B) uses a ratio $P(N|T_0)/P(N|T)$
that is the inverse of what you would expect from analogous
epidemiology studies. 
This is deliberate, and is shown to be necessary to ensure that the
expected number of claims attributed to climate change will equal the
expected change in the number of claims. 
The recorded number of lightning strikes are not determined
solely by the temperature but also have a clear seasonal dependence. 
To partially account for this, the dependence of lightning strikes on
temperature was assessed within each individual month from January
through to December.
This led to an estimated average increase in claims per month of 5.6\%
per degree Celcius. 
It was also possible to use the monthly 20th century average
temperatures to estimate how the number of strikes in 1996-2014 have
increased from the 20th century average, finding a 5.1\% increase 
per yearly-averaged temperature rise (in
degrees Celcius). 
The increases are much less than the typical 29\% variation in the
observed number of strikes per year. 
If US temperatures continue to increase at 0.19 degrees Celcius per
decade, then within ten years the estimates predict claims to increase
from those expected for average 20th century temperatures by  
4.8\% and 4.4\% respectively, with estimated increases in cost (for
the claims reported in the NCDC storm events database), of roughly
\$2.1M and \$1.9M  (not adjusted for inflation). 
The lower estimate is the same as would be attributed to climate
change by scheme A, and more comprehensively accounts for
non-uniform changes in average monthly temperatures (as opposed to
reporting a uniform average response), and therefore  
it seems likely to be a more accurate estimate. 
Scheme B attributes a 6.6\% increase in strikes per degree Celcius to
climate change, which is higher than scheme A, but over a sufficiently
long time period estimates A and B will converge to the same value. 
All the estimates here are less than the estimate of 12$\pm$5\%
increase in strikes per increase in average temperature by one degree
Celcius that is reported in \cite{Romps}. 
The cause of this difference unclear.
It could be caused by limitations in the data-driven analysis, such as
the use of monthly-averaged data for example.  
We end with a strong caveat originating from figure
\ref{YearAv}.
Figure \ref{YearAv} suggests that there may have been a qualitative
change in the climate in the past three years. 
It is possible that this is a naturally occurring change, or that it is
due to a change in the reporting of strikes, in which case it does not
reflect changes in climate and could be ignored. 
However, if it reflects a qualitative change in the climate then it is
possible that the estimates here could become irrelevant for future
studies. 
Because of Taylor's theorem \cite{Arfken}, a small linear response of
(statistically averaged) climate properties would be expected in
response to small changes in temperature, but if there is a strong
non-linear response then data-based studies such as the one described
here will become unreliable and climate modeling will be required.  
Nonetheless, studies such as the one here provide a useful baseline
from which to compare future behaviour, and allow a discussion of how 
best to quantitatively estimate and attribute the cost of climate
change.  

\vspace{.5cm}
{\bf Acknowledgments}
\vspace{.5cm}

Richard Clarke observed the exponential relationship between lightning
strikes and temperature that triggered this work and stimulated a
number of interesting discussions, some of which are expanded on in 
Refs. \cite{WebsterClarke, Clarke}.  
Thanks to Katie Webster for commenting on an earlier draft of this
paper.  

\end{document}